\documentclass[pss,fleqn]{w-art}
\usepackage{times}
\usepackage{w-thm}
\usepackage[]{graphicx}
\begin{document}
\DOIsuffix{theDOIsuffix}
\Volume{XX}
\Issue{1}
\Month{01}
\Year{2003}
\pagespan{1}{}
\Receiveddate{\sf } 
\Reviseddate{\sf  } 
\Accepteddate{\sf  } 
\Dateposted{\sf  }
\keywords{stripe phase, spiral phase, band structure, Hubbard model, 
doped cuprates.}
\subjclass[pacs]{71.10.Fd, 71.27.+a, 74.72.-h}



\title[Effect of the next-nearest neighbor hopping]
{Effect of the next-nearest neighbor hopping  on the stability and 
band structure of the incommensurate phases in the cuprates}


\author[M. Raczkowski] {Marcin Raczkowski\footnote{Corresponding
     author: e-mail: {\sf M.Raczkowski@if.uj.edu.pl},
     Phone: +48\,12\,663\,5628,
     Fax: +48\,12\,633\,4079}\inst{1}}
\address[\inst{1}] {Marian Smoluchowski Institute of Physics, Jagellonian
 University, Reymonta 4, 30059 Krak\'ow, Poland}

\author[R. Fr\'esard]{Raymond Fr\'esard\inst{2}}
\address[\inst{2}] {Laboratoire CRISMAT, UMR CNRS-ENSICAEN  6508, 
6 Bld. du Mar\'echal Juin, 14050 Caen, France}

\author[A.~M. Ole\'s]{Andrzej M. Ole\'s\inst{1,3}}
\address[\inst{3}] {Max-Planck-Institut f\"ur Festk\"orperforschung,
              Heisenbergstrasse 1, 70569 Stuttgart, Germany}

\begin{abstract}
Using a spin-rotation invariant version of the slave-boson 
approach we investigate the relative stability and band structure of various 
incommensurate phases in the cuprates.  
Our findings obtained in the Hubbard model with next-nearest neighbor 
hopping $-t'/t\simeq 0.15$, as appropriate for the La$_{2-x}$Sr$_x$CuO$_4$ 
family, support the formation of diagonal (vertical) stripe phases 
in the doping regime $x=1/16$ ($x=1/8$), respectively. In contrast, 
based on the fact that a larger value $-t'/t=0.3$ expected for 
YBa$_2$Cu$_3$O$_{6+\delta}$ triggers a crossover to the diagonal (1,1) 
spiral phase at increasing doping, we argue that it might explain 
why the static charge order has been detected in YBa$_2$Cu$_3$O$_{6+\delta}$ 
only in the highly underdoped regime.

\end{abstract}
\maketitle                   

The abundance of experimental results and theoretical investigations 
in recent years have shown that the ground state of the high-$T_c$ cuprates 
might be spatially inhomogeneous \cite{Kiv03}. In particular, both 
charge and spin modulations have been detected in neutron scattering studies 
performed on La$_{1.6-x}$Nd$_{0.4}$Sr$_x$CuO$_4$. 
Remarkably, incommensurate (IC) Bragg peaks indicate that in the lightly doped 
regime $x<1/16$ the antiferromagnetic (AF) domains are separated by 
\emph{diagonal} charge domain walls (DWs) \cite{Wak01} but at a higher 
doping level $x=1/8$ the doped holes self-organize into stripes that run 
\emph{vertically} across the CuO$_2$ planes \cite{Tra95}. On the one hand, 
the former are on average filled  by $1/\sqrt{2}$ hole per two DWs or even 
by one hole per one atom in a DW as established in La$_{2-x}$Sr$_x$CuO$_4$ 
(LSCO) around $x=0.02$ \cite{Fuj02}, which corresponds to the so-called 
\emph{filled} stripes. On the other hand, the latter are characterized by 
the filling of one hole per two atoms along the DWs meaning that they 
are only \emph{half-filled}. IC charge order 
consistent with vertical/horizontal stripes has also been  reported   
in highly underdoped YBa$_2$Cu$_3$O$_{6+\delta}$ (YBCO) with $\delta=0.35$ 
\cite{Moo02}, but despite several attempts no static charge order could have 
been detected at higher doping. Apart from explaining 
the neutron scattering data, stripe phases also offer a framework for 
interpreting a broad range of other experimental results such as 
angle-resolved photoemission spectroscopy (ARPES) \cite{Dam03} (see below) 
as well as magnetic excitation spectra in both LSCO and YBCO families that 
might be consistently explained in terms of fluctuating stripes 
\cite{Sei05,Voj06}.

While the stripe debate continues, an alternative scenario which might 
account for the IC spin structure in LSCO is a deformation of the AF 
order so as to optimize the hole motion that stabilizes a spiral phase 
\cite{Fre91,Has04}. Moreover, Lindg\aa rd \cite{Lin05} has shown that spiral 
states can also resolve the universality of magnetic excitations in the 
cuprates and provide a competing paradigm with the stripe phase concept.
Therefore, the main purpose of this paper is to address in a systematic way 
the competition between two possible \emph{site-centered} 
(i.e., with a fully suppressed magnetization along the domain walls) 
stripe ground states: 
half-filled vertical site-centered (HVSC) and filled diagonal 
site-centered (FDSC), their \emph{bond-centered} counterparts (where DWs 
are given by ladders with a weak ferromagnetic order on the rungs): 
half-filled vertical bond-centered (HVBC) and filled diagonal bond-centered 
(FDBC) DWs, as well as between the vertical (0,1) and diagonal (1,1) spiral 
phases. In order to treat noncanted stripe phases and the spiral order 
on equal footing and to implement local electron correlations we employ 
a rotationally invariant version of the slave-boson (SB) approach in spin 
space \cite{Zim97} by introducing auxiliary boson operators 
$\{e_i,d_i,p_{i0},\textbf{p}_i\}$ which control the actual electronic 
configuration at each site $i$. The Hamiltonian may be then written as,
\begin{equation}
H=-\sum_{ij}\sum_{\sigma\lambda\tau}t_{ij}
   z^{\dag}_{i\sigma\lambda}f^{\dag}_{i\lambda} 
   f^{}_{j\tau}z^{}_{j\tau\sigma} 
  + U\sum_{i}d^{\dag}_{i}d^{}_{i},
\label{eq:Hubb_sb}
\end{equation}
where $\{\underline{z}_i,\underline{z}_j\}$ are $2\times 2$ matrices in 
spin space which depend on the actual configuration of the boson fields, 
the hopping $t_{ij}$ is $t$ on the bonds connecting nearest-neighbor
sites and $t'$ for next-nearest neighbor sites, while $U$ is the on-site 
Coulomb interaction. Calculations were carried out on square 128$\times$128 
clusters which  became possible by developing an efficient scheme in 
reciprocal space which makes use of the stripe symmetry \cite{Rac06}. 
In our studies, we have chosen $U=12t$, which gives the ratio of $J/t=1/3$ 
(with the superexchange $J=4t^2/U$), being a value representative for 
LSCO \cite{Fle01}.

\begin{figure}[t!]
\begin{center}
\includegraphics*[width=0.9\textwidth]{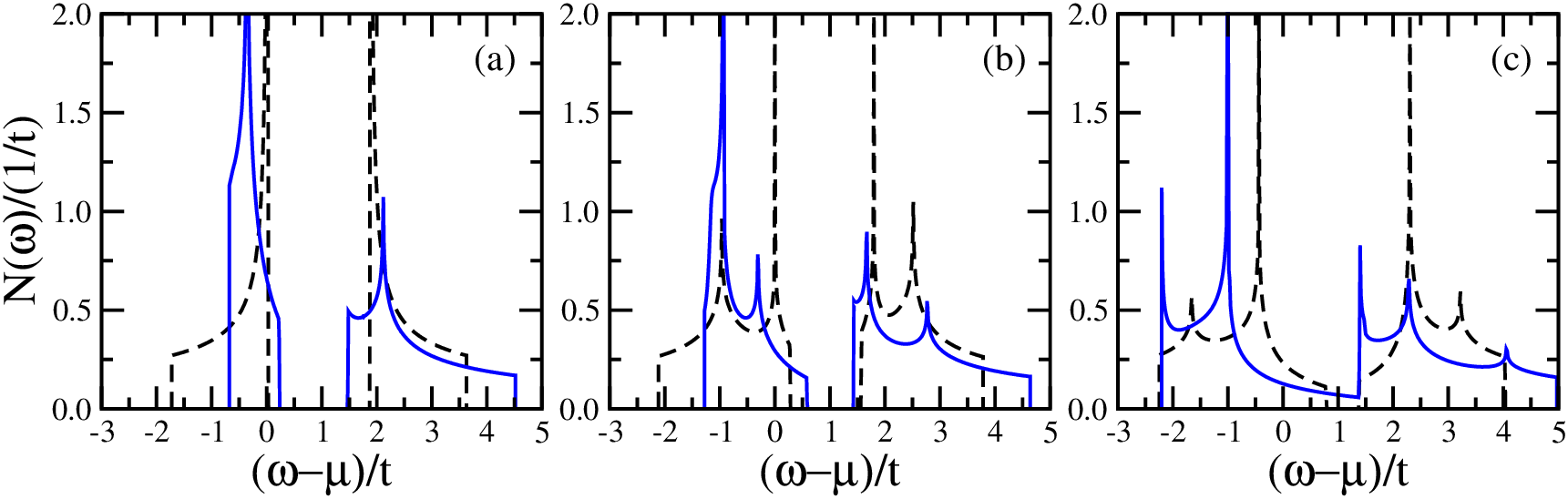}
\vspace*{-3mm}
\end{center}
\caption
{
Effect of the next-nearest neighbor hopping $t'$ on the density 
of states at doping $x=1/8$ as obtained in the extended Hubbard model 
for the uniform: 
(a) AF phase; 
(b) (0,1) spiral phase, and 
(c) (1,1) spiral phase. 
Parameters: $U=12t$, $t'=0$ (dashed line), and  $t'=-0.3t$ (solid line).  
}
\label{DOS}
\end{figure}
\begin{table}[b!]
\caption {
Local magnetization $m_i$ and double occupancy $d_i$ as found for 
the uniform AF state as well as for (0,1) and (1,1) spiral phases 
in the extended Hubbard  model with $U=12t$ for $x=1/16$, and  $x=1/8$.  
}
\label{mds}
\begin{tabular}{ccccccccccc}
\hline
\multicolumn{2}{c}{}              &
\multicolumn{4}{c}{$x=1/16$}      &
\multicolumn{1}{c}{}              &
\multicolumn{4}{c}{$x=1/8$}       \\
\multicolumn{2}{c}{}              &
\multicolumn{2}{c}{$t'=0$}        &
\multicolumn{2}{c}{$t'/t=-0.3$}   &
\multicolumn{1}{c}{}              &
\multicolumn{2}{c}{$t'=0$}        &
\multicolumn{2}{c}{$t'/t=-0.3$}   \\
\hline
phase      & & $m_i$  & $d_i$  &  $m_i$  & $d_i$  &  
           & $m_i$  & $d_i$  &  $m_i$  & $d_i$ \\
\hline
 AF        & & 0.803  & 0.0265 &  0.799  & 0.0257 & 
           & 0.586  & 0.0227 &  0.595  & 0.0207 \\
(0,1)      & & 0.813  & 0.0257 &  0.811  & 0.0247 & 
           & 0.646  & 0.0216 &  0.656  & 0.0190 \\
(1,1)      & & 0.820  & 0.0251 &  0.829  & 0.0229 & 
           & 0.682  & 0.0202 &  0.730  & 0.0158 \\
\hline
\end{tabular}
\end{table}

Regarding the excitation spectrum of the Hubbard model, dynamical 
mean-field theory (DMFT) revealed that it consists of a lower Hubbard band 
(LHB), an upper Hubbard band (UHB), and a quasiparticle peak that vanishes at 
the Mott transition \cite{Kot96}.
Since the former two are dynamical in nature, they are hardly present 
in mean-field calculations, while the latter is at the heart of 
the Brinkman-Rice description of the Mott transition and of the equivalent 
SB mean-field approach in the paramagnetic phase.
In contrast, in stripe phases bands, separated by energies of order $U$ 
and corresponding to the LHB and UHB, are part of the spectrum, 
on top of midgap bands in which the chemical potential may be located, 
as discussed below.

Let us now review the mechanisms responsible for the stability of these 
various IC phases in the pure Hubbard model with $t'=0$ at moderate doping 
$x<0.15$. First of all, Fig.~\ref{DOS} shows that a prominent feature 
of the spiral phases is the van Hove singularity below the Fermi energy 
that leads to a substantial energy gain with respect to the AF phase. 
Remarkably, this energy gain is amplified by doping making the spiral 
phases increasingly competitive. Next, as listed in Table~\ref{mds}, 
they are characterized by larger staggered magnetization resulting into 
smaller double occupancy and consequently into a better Coulomb energy 
as compared to the AF phase.
Finally, one finds that this mechanism is more effective when the spiral 
twist is along the diagonal of the Brillouin zone.       
In contrast, stripe phases result from a compromise between the optimized 
kinetic energy of doped holes and the superexchange interaction \cite{Rac06}. 
Indeed, propagation of doped holes is easy when the magnetization is 
locally suppressed along the charge DWs, while the superexchange 
is best optimized when the AF domains are close to half-filling.

\begin{table}[t!]
\caption {
Free energy $F/t$ per site for various phases at $x=1/8$ found in the SB 
approximation in the extended Hubbard model with $U=12t$ for representative 
values of the next-nearest neighbor hopping: 
$t'=0$, $t'=-0.15t$, and $t'=-0.3t$. 
The most advantageous energy for each $t'$ is given in bold characters. 
}
\label{1_8}
\begin{tabular}{rccccccc}
\hline
$t'/t$  &      AF    &    (0,1)   &  (1,1)     &    FDSC     &     FDBC  
        &    HVSC    &    HVBC    \\
\hline
 0.00   &  $-$0.5393 &  $-$0.5613 &  $-$0.5700 & {\bf $-$0.5821}  &  $-$0.5819
        &  $-$0.5689 &  $-$0.5680 \\
     
$-$0.15 &  $-$0.5323 &  $-$0.5564 &  $-$0.5699 & {\bf $-$0.5716}  &  $-$0.5713 
        &  $-$0.5700 &  $-$0.5691 \\
       
$-$0.30 &  $-$0.5341 &  $-$0.5594 &  {\bf $-$0.5842} &  $-$0.5655 &  $-$0.5651
        &  $-$0.5749 &  $-$0.5740  \\ 
\hline
\end{tabular}
\end{table}
\begin{table}[b!]
\caption {
The same as in Table \ref{1_8} but for $x=1/16$. FDBC stripe phase 
is unstable at $t'=-0.3t$. 
 }
\label{1_16}
\begin{tabular}{rccccccc}
\hline
$t'/t$  &      AF    &    (0,1)   &  (1,1)     &    FDSC     &     FDBC  
        &    HVSC    &    HVBC    \\
\hline
 0.00   &  $-$0.4183 & $-$0.4286  &  $-$0.4363 & {\bf $-$0.4561}  &  $-$0.4560
        &  $-$0.4496 & $-$0.4492  \\
$-$0.15 &  $-$0.4173 & $-$0.4287  &  $-$0.4376 & {\bf $-$0.4508}  &  $-$0.4507
        &  $-$0.4503 & $-$0.4499  \\
$-$0.30 &  $-$0.4194 & $-$0.4319  &  $-$0.4444 &    $-$0.4478 &    -- 
        & {\bf $-$0.4529} & $-$0.4525  \\  
\hline
\end{tabular}
\end{table}

Turning now to the extended Hubbard model, the main effect of $t'$ 
is to shift the van Hove singularity towards negative energy \cite{Fle97}. 
As a result, perfect nesting of the Fermi surface 
is broken and a metal-to-insulator transition occurs for a finite $U_c$, 
even at half-filling \cite{Kot00}. Accordingly, antiferromagnetism is 
suppressed at $x=0$ becoming, however, enhanced when moving towards the 
van Hove point.  
Similarly, the filled stripe phases should not benefit from $t'$ 
as it strongly frustrates the antiferromagnetism in the weakly doped 
AF domains separating DWs.  In contrast, one can conjecture that the spiral 
phases better make use of the van Hove singularity by its systematic shift 
below the Fermi energy. This, together with both larger staggered 
magnetization and smaller double occupancy, should yield a more significant 
energy gain and make the spiral phases more competitive with increasing 
$-t'/t$.  Therefore, one may expect a critical $t'$  for which the (1,1) 
spiral phase becomes the ground state. 
As shown in Table~\ref{1_8}, this indeed happens for $t'\simeq -0.2t$ 
at doping $x=1/8$. In this case, one observes a spectacular transfer of 
density of states from the Fermi energy towards the bottom of the LHB 
as depicted in Fig.~\ref{DOS}(c). 
In addition, staggered magnetization in the (1,1) spiral phase is strongly 
enhanced as compared to the AF phase supported by concomitant reduction 
of double occupancy (see Table~\ref{mds}). However, Table~\ref{1_16} shows 
that for small doping $x=1/16$, the above mechanisms are not sufficient 
to stabilize the spiral order even for $t'/t=-0.3$.

\begin{figure}[t!]
\begin{minipage}[t]{0.6\textwidth}
\includegraphics*[width=\textwidth]{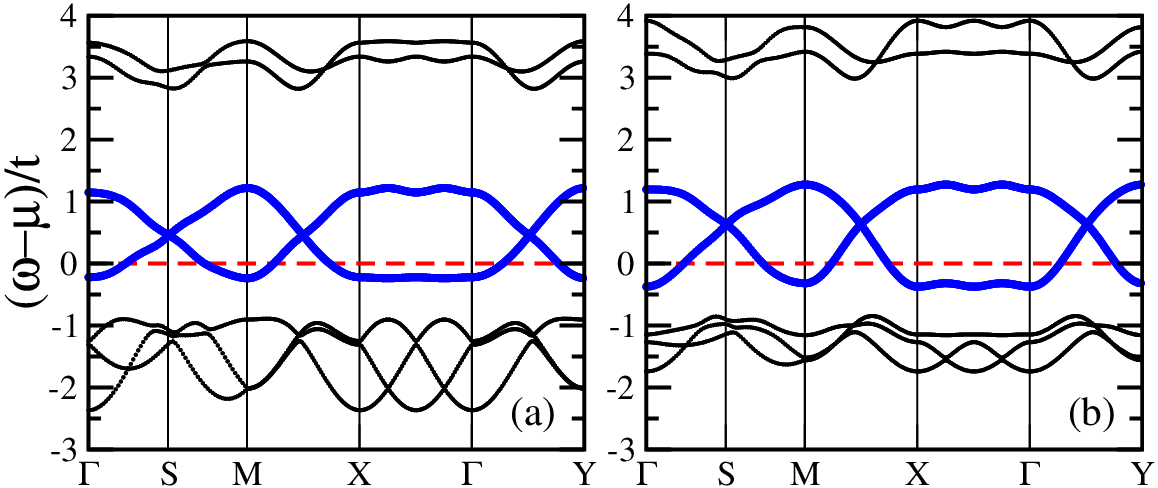}
\caption
{
Low energy part of the electronic structure found at \mbox{$x=1/8$} 
for the HVSC stripe phase in the extended Hubbard model with \mbox{$U=12t$} 
as well as with: (a) $t'=0$ and (b) $t'=-0.15t$. 
Thick solid line depicts midgap states while dashed line indicates 
the Fermi level. The highest energy band is centered around 
$\omega -\mu \simeq 7t$.
}
\label{HFSC}
\end{minipage}
\hfil
\begin{minipage}[t]{0.31\textwidth}
\includegraphics*[width=\textwidth]{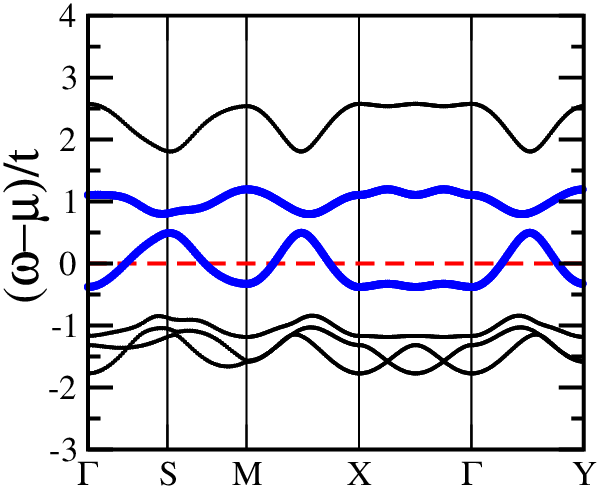}
\caption
{
The same as in Fig.~\ref{HFSC}(b) but for the HVBC stripe phase. 
The two highest energy bands are centered around 
$\omega -\mu \simeq 6t$.
}
\label{HFBC}
\end{minipage}
\end{figure}

Remarkably, since the holes are more delocalized over the AF domains 
which are then less frustrated, finite $t'$ also promotes the half-filled 
vertical stripe phase. In addition, it induces conspicuous changes in the band 
structure that should have measurable consequences in ARPES experiments. 
First of all, as shown in Fig.~\ref{HFSC}(a) segregation of doped holes into 
\emph{metallic} site-centered DWs gives rise to the appearance of two 
additional \emph{partially} filled bands lying within the Mott-Hubbard gap. 
Secondly, as a consequence of the vertical stripe order, 
the $\Gamma-X$ and $\Gamma-Y$ directions with $X=(\pi,0)$ and $Y=(0,\pi)$ 
are inequivalent. Indeed, as the fully quenched spin polarization 
facilitates the propagation of the holes along the site-centered DWs, 
the maximum dispersion is expected along the $\Gamma-Y$ direction. 
In contrast, the midgap bands almost do not disperse 
along the $\Gamma-X$ direction, which indicates that the holes are rather 
static along the horizontal direction.  Therefore, both midgap bands could 
be approximated by one-dimensional noninteracting dispersion  
$\varepsilon_{\textbf{k}}^{\textrm{DW}} =\pm 2t\cos k_y + \textrm{cst}$.
It crosses the Fermi energy $\mu$ at $\textbf{k}=(\pi,\pi/4)$, 
$(\pi/4,\pi/4)$, and at the equivalent points as expected for the stripes 
separated by four lattice spacings. However, due to strong local correlations 
the bandwidth of these states is strongly renormalized and reduced 
down to $W_{\textrm{DW}}/t\simeq 1.45$. 
On the one hand, some features depicted in Fig.~\ref{HFSC}(a) such 
as the flat band near the $X$ point crossing the Fermi level at the 
$\textbf{k}=(\pi,\pi/4)$ point and excitations at the $S=(\pi/2,\pi/2)$ 
point at a higher binding energy that at the $X$ point, are in a close 
agreement with the ARPES spectral density for LSCO at $x=1/8$ \cite{Dam03}. 
On the other hand, no spectral weight was found at the Fermi energy 
along the nodal $\Gamma-M$ direction with $M=(\pi,\pi)$, 
especially at the $(\pi/4,\pi/4)$ point. Therefore, one has to use more 
accurate methods, such as the DMFT, which include dynamical correlations  
and leads to a better agreement with the experimental data, reproducing a
pseudogap at $x=1/8$ \cite{Fle01}. 
Furthermore, Fig.~\ref{HFSC}(b) indicates that a negative $t'=-0.15t$, as 
expected for LSCO, leads to a distinct broadening of the midgap bandwidth up 
to $W_{\textrm{DW}}/t\simeq 1.6$. Consequently, it shifts these states 
to a lower energy which explains the increasing stability of the half-filled 
DWs in spite of the concomitant narrowing of the LHB bandwidth associated with 
the insulating AF background.   
Finally, similar conclusions might be drawn concerning the effect of $t'$ on 
the band structure of the HVBC stripe phase shown in Fig.~\ref{HFBC}. 
In this case, one finds a gap between both midgap states 
due to a finite AF spin polarization along the bond-centered DWs. However, 
shape of the band structure depends on the charge periodicity. Indeed,  
\emph{even} period stripes yield a spectral gap at the $S$ point while this 
gap is found to be absent for \emph{odd} period ones \cite{Gra06}.

\begin{figure}[t!]
\begin{minipage}[t]{0.6\textwidth}
\includegraphics*[width=\textwidth]{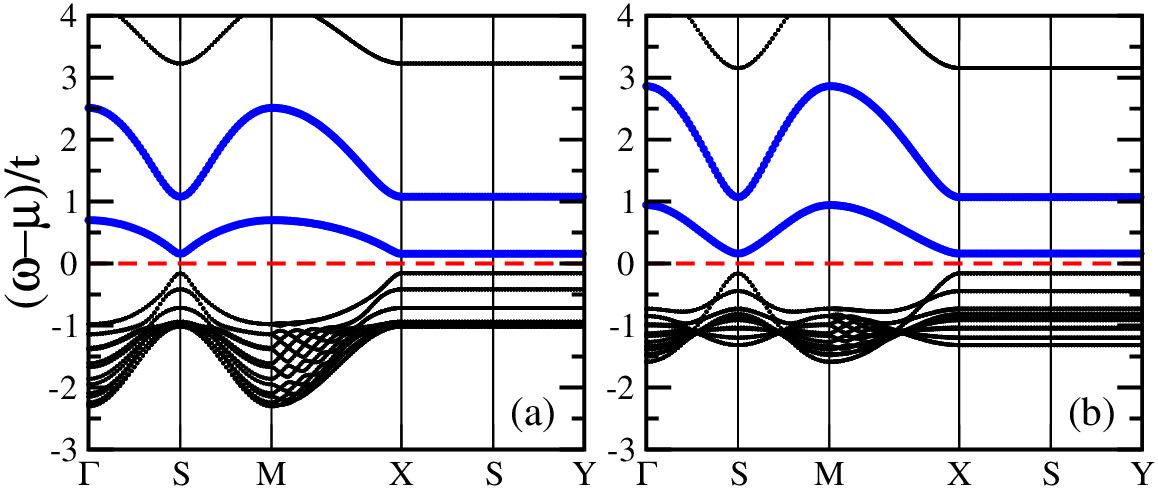}
\caption
{
Low energy part of the electronic structure found at \mbox{$x=1/16$} 
for the FDSC stripe phase in the extended Hubbard model with \mbox{$U=12t$} 
as well as with: (a) $t'=0$ and (b) $t'=-0.15t$. The highest energy bands (not
shown) are centered around $\omega -\mu \simeq 9t$.
}
\label{FSC}
\end{minipage}
\hfil
\begin{minipage}[t]{0.31\textwidth}
\includegraphics*[width=\textwidth]{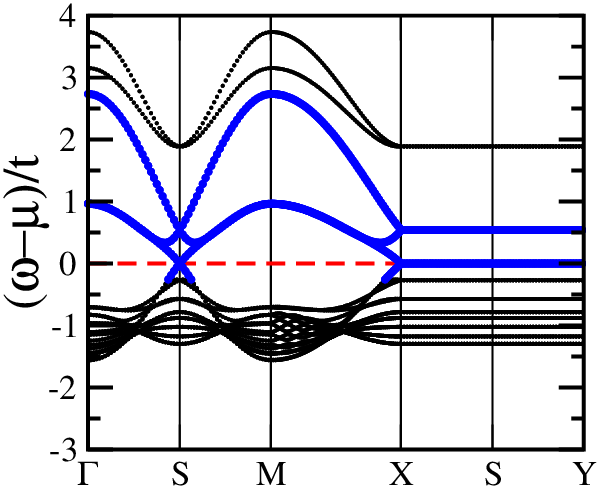}
\caption
{
The same as in Fig.~\ref{FSC}(b) but for the FDBC stripe phase. The highest
energy bands (not shown) are centered around $\omega -\mu \simeq 9t$. 
}
\label{FBC}
\end{minipage}
\end{figure}

An entirely different mechanism is responsible for the stability of 
the \emph{filled} site-centered DWs. It might be best understood from the 
band structure shown for $x=1/16$ in Fig.~\ref{FSC}(a). Here, 
each diagonal DW induces the formation of two (dispersionless along 
the $X-Y$ direction) entirely \emph{unoccupied} midgap bands. 
Thus the special stability of these phases follows from a real gap
that opens  
in the symmetry broken state between the highest occupied state of the LHB 
and the bottom of the midgap band \cite{Zaa96,Ich99}. 
Besides, the main effect of increasing $-t'/t$ is to narrow the LHB and to
push its lowest energy  
states towards higher energy while the midgap states remain intact 
as depicted in Fig.~\ref{FSC}(b). 
This feature is clearly visible along the parallel to the stripes 
$\Gamma-M$ direction and strongly decreases the stability 
of the FDSC stripe phase with respect to the half-filled one 
(see Tables~\ref{1_8} and \ref{1_16}). Unfortunately, the above gap is 
inconsistent with the ARPES data on the lightly doped LSCO. Indeed, 
experimentally a quasiparticle peak crosses the Fermi level in the nodal 
direction $\Gamma-M$ and forms a hole pocket at the $S$ point \cite{Yos03}. 
A possible interpretation is provided by Fig.~\ref{FBC} showing that finite 
spectral weight around the $S$ point may arise from the bond-centered DWs. 
Here, finite spin polarization of DWs leads to the hybridization 
of the midgap bands with the LHB. Such feature was also found in the two-band 
Hubbard model describing stripe phases in the nickelates \cite{deg}.

In conclusion, we have found that the next-nearest neighbor hopping $t'$ 
plays an important role in affecting the relative stability and the 
band structure of the IC phases. 
Next, partial filling of the bands within the Mott-Hubbard gap, pseudogap 
that forms at the Fermi energy around the $S$ point, and the flat band 
near the $X$ point crossing the Fermi level at the $\textbf{k}=(\pi,\pi/4)$ 
point are found to be general consequences of the half-filled vertical 
stripe phase in qualitative agreement with the ARPES results established 
at doping $x=1/8$. 
In contrast, finite spectral weight around the $S$ point experimentally
observed at $x=1/16$ doping might be attributed to the FDBC stripe phase 
also explaining the metallic behavior of the lightly doped LSCO. 
Finally, we would like to emphasize that the strongly enhanced stability of
the (1,1) spiral phase by finite $t'$ above $x=1/8$ with respect to 
the above stripe phases is robust and is not 
affected even by optimizing the stripe filling that further 
lowers the energy of the stripe phases \cite{unpub}. This may explain why 
the static charge order has been detected in YBa$_2$Cu$_3$O$_{6+\delta}$ 
only in the highly underdoped regime.

\begin{acknowledgement}
We thank M. Granath for helpful comments.
This work was supported by the Polish Ministry of Science and Education  
under Project No. 1~P03B~068~26 and by the Minist\`ere
Fran\c{c}ais des Affaires Etrang\`eres under POLONIUM 09294VH.
\end{acknowledgement}


\end{document}